\documentclass[journal=jacsat,manuscript=article]{achemso}

\usepackage[utf8]{inputenc}
\usepackage[T1]{fontenc}
\usepackage{mathptmx}
\usepackage{xcolor}
\usepackage[version=3]{mhchem} 

\author{João Victor Lemos Vale}
\affiliation[UFBA]{Instituto de F{\'i}sica, Universidade Federal da Bahia, Campus Universit{\'a}rio de Ondina, Salvador 40210-340, BA, Brazil}

\author{Lucas Cesena}
\affiliation[UFBA]{Instituto de F{\'i}sica, Universidade Federal da Bahia, Campus Universit{\'a}rio de Ondina, Salvador 40210-340, BA, Brazil}

\author{Bruno H. S. Mendon\c{c}a}
\affiliation{Departamento de F{\'i}sica, ICEX, Universidade Federal de Minas Gerais, CP 702, Belo Horizonte 30123-970, MG, Brazil}

\author{Elizane E. de Moraes}
\email{elizane.fisica@gmail.com}
\affiliation[UFBA]{Instituto de F{\'i}sica, Universidade Federal da Bahia, Campus Universit{\'a}rio de Ondina, Salvador 40210-340, BA, Brazil}
\affiliation{Catalan Institute of Nanoscience and Nanotechnology (ICN2), CSIC and BIST, Campus UAB, Bellaterra, 08193 Barcelona, Spain}
\title[An \textsf{achemso} demo]
  {Dipole Alignment and Layered Flow Structure in Pressure-Driven Water Transport through MoS$_{2}$ Membranes}

\abbreviations{IR,NMR,UV}

\begin{document}

\begin{abstract}
Efficient water transport through nanostructure membranes is essential for advancing filtration and desalination technologies. In this study, we investigate the flow of water through molybdenum disulfide (MoS$_{2}$) nanopores of varying diameters using molecular dynamics simulations. The results demonstrate that both pore size and atomic edge composition play crucial roles in regulating water flux, molecular organization, and dipole orientation. Larger pores facilitate the formation of layered water structures and promote edge-accelerated flow, driven by strong electrostatic interactions between water molecules and exposed molybdenum atoms. In narrower pores, confinement and asymmetric edge chemistry induce the ordered alignment of dipoles, thereby enhancing directional transport. Velocity and density maps reveal that pore edges act as active zones, concentrating flow and reducing resistance. These findings highlight the significance of pore geometry, surface chemistry, and molecular dynamics in influencing water behavior within MoS$_{2}$ membranes, providing valuable insights for the design of advanced nanofluidic and water purification systems.
\end{abstract}

\section{Introduction} 
\label{sec:introduction}

Access to clean water is becoming increasingly difficult in many parts of the world. As climate change intensifies and populations grow, ensuring safe and sustainable water supplies has become one of the most urgent challenges of our time. In response, researchers have turned to innovative technologies like nanofiltration and desalination—especially those that use two-dimensional (2D) materials known for their remarkable structural and chemical properties \cite{suk2010water,cohen2012water,@10.1021/nl200843g,10.1080/1536383X.2024.2445668,abal2021molecular,yasmeen2025enhanced,10.1016/j.cej.2024.158366,10.1017/flo.2024.34,gao2017rational,10.1039/D4CP01068J,10.1126/sciadv.adj3760,10.1002/smll.202409950,10.1063/1.4893782,cheng2025research,vo2024recent}. Graphene was once the star of this field, praised for its strength and thinness. But its chemical stability and the difficulty of creating ultra-small pores have limited its practical use in large-scale water purification. That's where molybdenum disulfide (MoS$_{2}$) — an inorganic compound of molybdenum (Mo) and sulfur (S) from the transition metal dichalcogenide (TMD) family — comes into the picture. MoS$_{2}$ stands out for its semiconducting nature, mechanical durability, and ability to be chemically modified, making it a strong candidate for next-generation water filtration membranes \cite{qasim2019reverse,10.1038/ncomms9616,10.1021@jp205877b,sun2025water,suk2010water,li2023modulation,abal2021molecular,@10.1021/acs.jpcb.3c02889,@10.1002/aic.17543,li2024molecular,@10.1002/aic.17543,10.1016@j.carbon.2017.12.039,10.1103/PhysRevLett.102.184502,kargar2019water,zhou2024interlink,li2024electric,yang2023fast,10.1103/PhysRevE.111.L023101}.

MoS$_{2}$ membranes are especially appealing due to the possibility of tailoring their pore structures at the atomic scale, allowing the selective transport of water molecules while rejecting ions and contaminants \cite{10.1038/ncomms9616,feng2015electrochemical,lee2012synthesis,10.1021/acsnano.5b01281,10.1126/science.1194975,10.1002/admi.202500604,romanov2024impact,smith2011large,barati2024fast}. Such control over molecular transport is critical for desalination, wastewater treatment, and nanofluidic devices. Recent experimental and theoretical studies have shown that MoS$_{2}$ nanopores can outperform conventional polymeric membranes in terms of both water permeability and salt rejection, highlighting their potential for energy-efficient water purification \cite{10.1038/ncomms9616,zhao2019nanoscale,mi2014graphene}.

A fundamental step toward realizing these applications is a deep understanding of water flow dynamics through MoS$_{2}$ membranes. Unlike bulk systems, transport at the nanoscale is governed not only by pore size and geometry but also by confinement effects, interfacial interactions, and the specific chemistry of the pore edges \cite{falk2010molecular,tocci2014friction}. Investigating these aspects experimentally is often limited by constraints on resolution and reproducibility. In this regard, computational simulations provide a powerful complementary approach, enabling atomistic insights into water transport mechanisms, hydrogen-bonding behavior, and the influence of surface functionalization under controlled conditions \cite{zhao2019nanoscale}.

Therefore, the study of water flow in MoS$_{2}$ membranes is not only crucial for optimizing their performance in desalination and filtration but also for advancing fundamental knowledge in nanofluidics. Such understanding can guide the rational design of new materials and devices for critical applications in sustainable water management, energy conversion, and bioengineering.

This study investigates the behavior of water molecules as they flow through MoS$_{2}$ nanopores with varying diameters, utilizing molecular dynamics simulations to capture the intricacies of this process. The findings reveal that both the size of the pore and the chemical nature of its atomic edges are key factors in shaping the dynamics of water transport. MoS$_{2}$ membranes display distinctive flow patterns, marked by accelerated movement near the pore edges and the emergence of layered molecular structures. These phenomena are closely linked to electrostatic interactions between water molecules and the exposed molybdenum and sulfur atoms, which influence how water organizes itself and navigates through the confined space. The remainder of this manuscript is organized as follows: In Sec. II, we present the simulation methodology used in this analysis and define the simulated models. In Sec. III, we discuss the results, and in Sec. IV, we present the conclusions.

\section{Methodology and Computational Details}

\begin{figure}[H]
    \centering
    \includegraphics[width=6.4in]{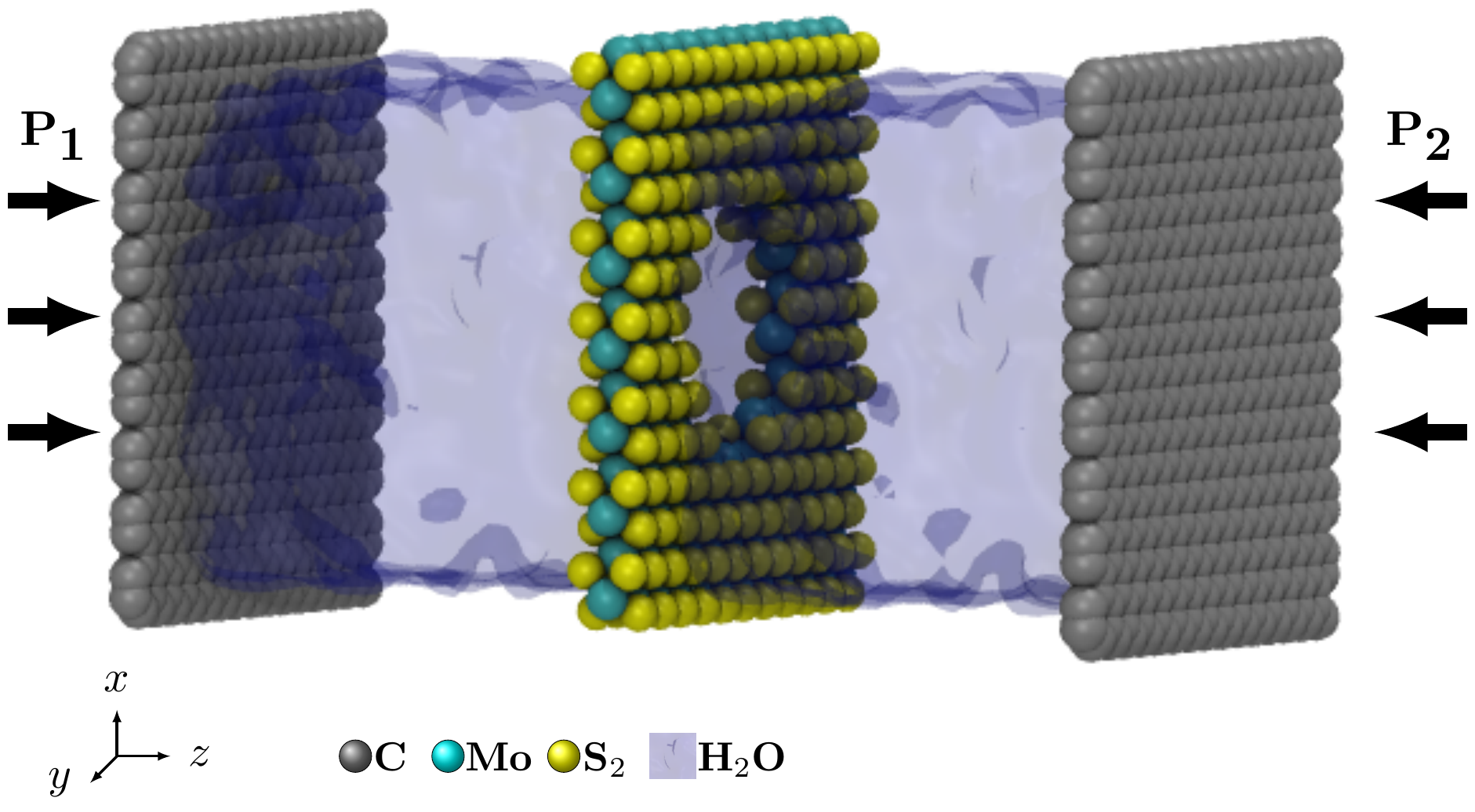}
    \caption{Side view of the simulation box. The y-axis is out-of-plane, towards the reader.}
    \label{fig:system}
\end{figure}

\begin{figure}[H]
    \centering
    \includegraphics[width=6.4in]{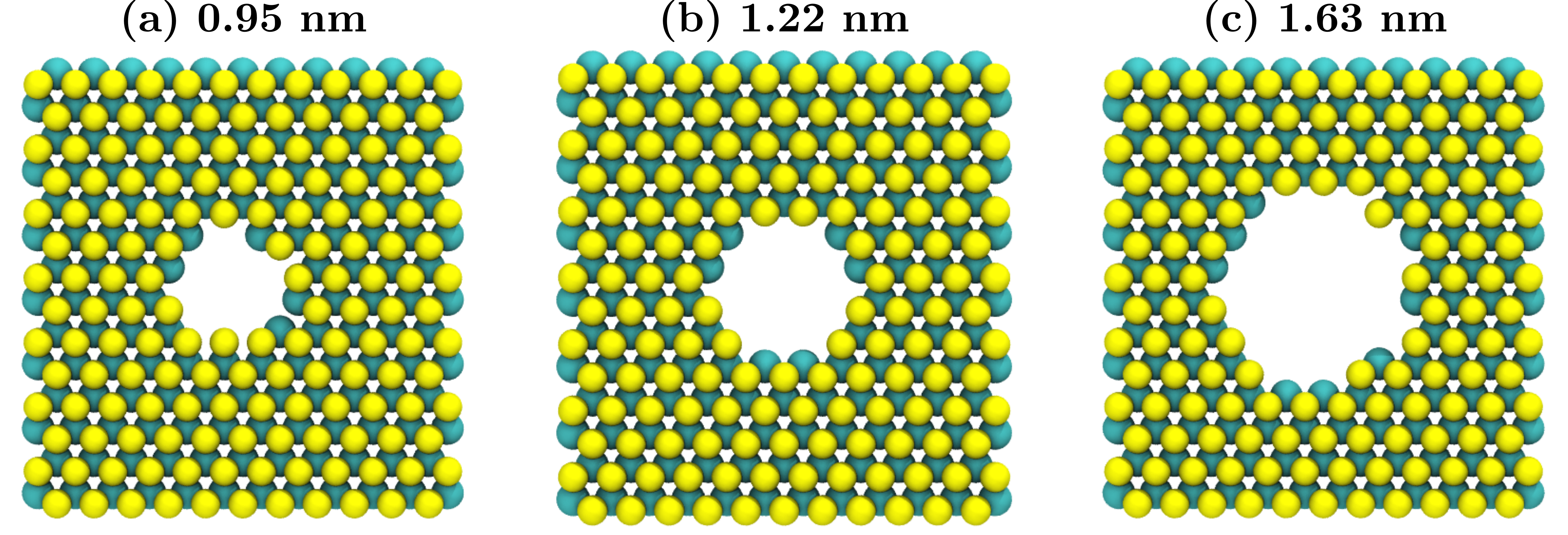}
    \caption{Nanoporous membranes structures with different pores diameters.(a)-(c) represents the represents the MoS$_{2}$ membrane.}
    \label{fig:pore}
\end{figure}

To investigate the flux of water molecules through MoS$_{2}$ membranes, we designed the simulation box as illustrated in Figure \ref{fig:system}. The cell dimensions are L$_{x}$ $\approx$ 36 \AA, L$_{y}$ $\approx$ 37 \AA, and L$_{z}$ = 70 \AA. Periodic boundary conditions were applied in all directions. The system contains two reservoirs filled with 1000 water molecules each, two graphene sheets (piston 1 and piston 2) located at the edges in the z-direction, and a 2D membrane perforated with a nanopore of variable size at its center, located at the origin of the box. We designed membranes with 3 different pore diameters, where the pore diameters are $0.95$ nm, $1.22$ nm, and $1.63$ nm. These values were obtained from the pore diameters of carbon nanotubes (CNT) (7,7), (9,9), and (12,12) (see Figure \ref{fig:pore}).

The membranes and pistons are treated with rigid bond lengths and angles, modeled by the Lennard-Jones potential (LJ) and, for the case of MoS$_{2}$, the Coulombic potential also. The water molecules are represented by the TIP4P-2005 force field~\cite{Abascal2005-iu}. In this water model, there is only one LJ interaction site placed at the oxygen position with parameters $\sigma_{OO}$ and $\epsilon_{OO}$. The positive charges $q_H$ are placed at the hydrogen positions, while the negative charge $q_M$ is placed at a virtual massless site M located at the H-O-H bisector, $d_{OM}$ distance units apart from the oxygen atom. The cross-interaction parameters between all the species are provided by Lorentz-Beherlort mixing rules. All the force field parameters are summarized in Table~\ref{tab_parameters}.

\begin{table}[H]
	\begin{center}
	\caption{Force field parameters utilized. The units of the LJ parameters are $\mathrm{kcal.mol^{-1}}$ for $\epsilon_{XX}$ and \AA \; for $\sigma_{XX}$, where $X=\mathrm{(C,O,H,Mo,S)}$. The charges are given in elementary charge (\textit{e}) units.}
		\begin{tabular}{ c c c c c c c }
			\hline
			   $\mathrm{H_2O}^{(a)}$ && $\mathrm{MoS_2}^{(b)}$ && $\mathrm{Gr}^{(c)}$ (Pistons) &  \\ \hline
              $\epsilon_{OO}$= 0.1852  && $\epsilon_{Mo}$= 0.0135 && $\epsilon_{CC}$= 0.086 & \\
              $\epsilon_{HH}$= 0.0000  && $\epsilon_{SS}$= 0.4612 && $\sigma_{CC}$= 3.40   & \\
              $\sigma_{OO}$= 3.1589    && $\sigma_{Mo}$= 4.200    && $q_C$= 0.00  & \\
              $\sigma_{HH}$= 0.0000    && $\sigma_{SS}$= 3.130    &&              & \\
              $q_M$= -1.1128           && $q_{Mo}$= 0.6           &&              & \\
              $q_H$= 0.5564            && $q_S$= -0.3             &&              & \\
              $r_{OH}$ (\AA)= 0.9572 &&                &&              & \\
              $\theta_{HOH}$ ($\mathrm{deg}$)= 104.52 &&             &&              & \\
              $d_{OM}$ (\AA)= 0.1546  &&               &&              & \\
            \hline 
		\end{tabular}
        \\
        \footnotesize{(a): Vega and Abascal~\cite{Abascal2005-iu}; (b): Liang et al.~\cite{10.1103/PhysRevB.79.245110}; (c): Hummer et al.~\cite{Hummer2001}}
		\label{tab_parameters}
	\end{center}
\end{table}

The simulations were performed by using the Large-scale Atomic/Molecular Massively Parallel Simulator (LAMMPS) software, with a timestep of 1.0 fs. The SHAKE algorithm was employed to maintain the bond lengths and angles constrained. Coulombic long-range interactions were computed by the PPPM solver with a precision of 10$^{-4}$. We used a cutoff of 12 \AA \; for both LJ and Coulombic interactions. The simulation protocol is given as follows:

\begin{enumerate}
    \item Pre-equilibrium in the NVE ensemble with a 0.1 ns  MD run to minimize system energy, keeping the pistons frozen (net force equal to zero);
    \item Equilibration in the NVT ensemble at 300 K during 0.2 ns Nosé-Hoover thermostat ;
    \item Forces are applied in the pistons in order to impose 1 bar in each system to reach the water equilibrium densities at 300 K. Equilibration in the NPT ensemble during 1.0 ns; 
    \item  Pistons are frozen in the new equilibrium position. The NVT ensemble's equilibrium at 300 K was controlled via the Nosé-Hoover thermostat during 1.0 ns. 
    \item Nanopores are opened. Different forces are applied in each piston to mimic the pressure gradient. NPT ensemble during 10 ns at 300 K and different feed pressures.
\end{enumerate}

The water flow occurs along the z-axis, normal to the membrane, pressure-driven by pistons 1 and 2, illustrated in Figure \ref{fig:system}. The pressure difference is introduced by applying an external force $F$ on each atom of the pistons in the z-direction, calculated according to Equation \ref{eq1}.

\begin{equation}
    F = \frac{P \cdot A}{n}
    \label{eq1}
\end{equation}

\noindent where $n$ is the number of atoms, $A$ is the surface area, and $P$ is the pressure on the surface. Piston 2 (right) has a fixed pressure value of 1 bar, while in piston 1 (left), the pressure was chosen so that the pressure gradients of 100 MPa, 200 MPa, 300 MPa, 400 MPa, and 500 MPa were achieved. 

The system's dynamic features were evaluated by considering the flow rate calculations as given by Equation \ref{eqflow}.

\begin{equation}
    \phi_{H_2O} = \lambda \cdot V_{mol} \cdot v
    \label{eqflow}
\end{equation}

\noindent where $\lambda$ ($\mathrm{molecules} \cdot \mu m^{-1}$) is the linear number density of water molecules, V$_{mol}$ $\mu m^3)$ is the average volume of a water molecule, and $v$ ($\mu m^3 \cdot s^{-1}$) is the water flow velocity acquired from the least-squares linear regression line fitted to the data cloud that relates the average molecular displacement as a function of time, as taken from the MD trajectory file. The calculations are performed inside a rectangular box with dimensions $\Delta x = L_x$, $\Delta y = L_y$, and $\Delta z = 4$ \AA \; centered at the pore.

\section{Results and discussion}

\begin{figure}[H]
\centering
\includegraphics[width=4.in]{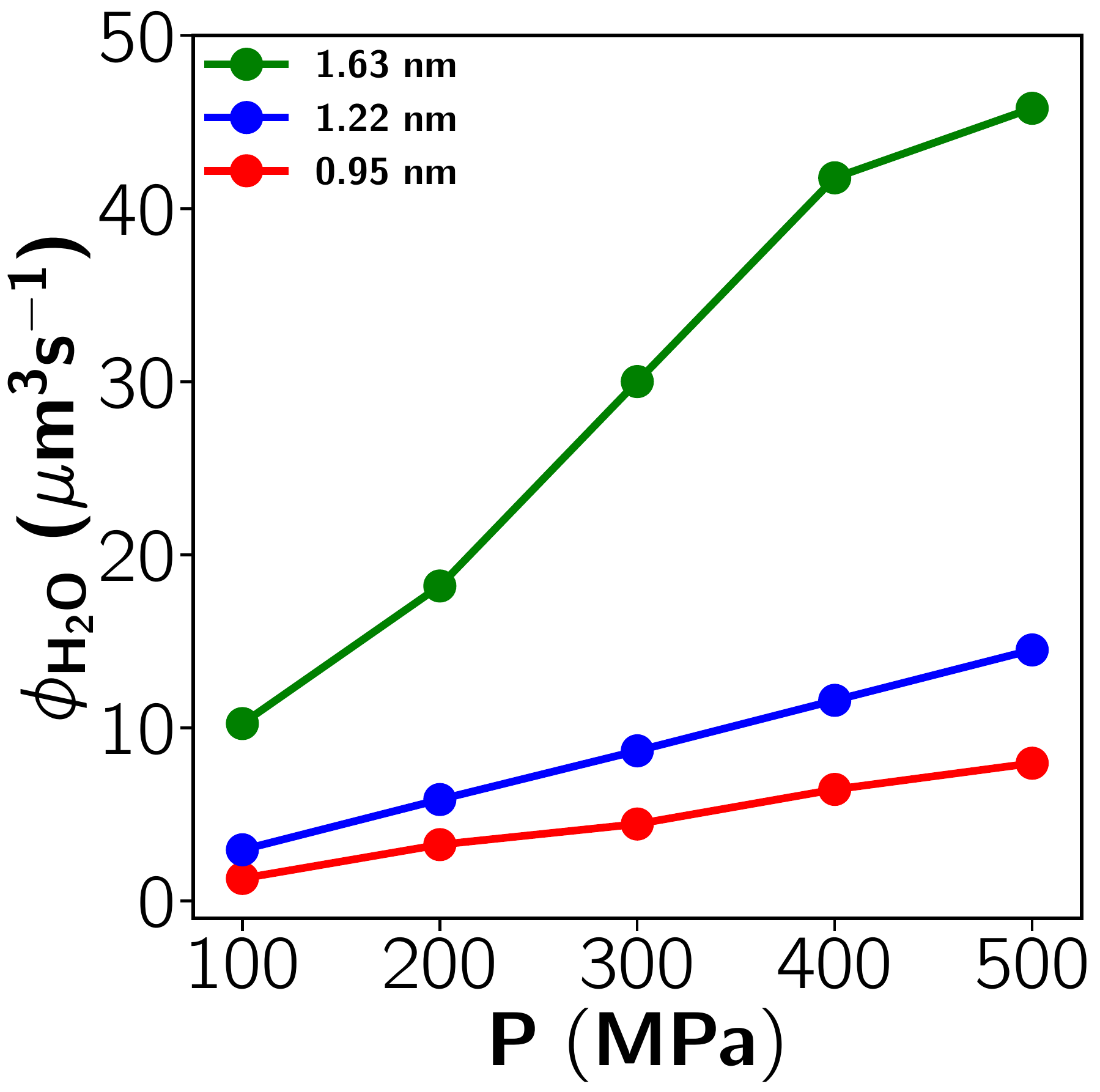}
\caption{Water flow rates through molybdenum disulfide nanopores at various diameters: 0.95 nm (red), 1.22 nm (blue), and 1.63 nm (green).}
\label{fig:flow1}
\end{figure}

Molecular dynamics simulations reveal that water flows through molybdenum disulfide (MoS$_{2}$) membranes significantly faster than graphene membranes of comparable pore diameters \cite{del2023ultrafast}.  The water flux observed in Fig.\ref{fig:flow1} demonstrates the influence of pore diameter and applied pressure on this transport. The increase in pore diameter in MoS$_{2}$ membranes is directly correlated with a significant increase in water flux. This phenomenon can be explained by the reduction in hydrodynamic resistance and changes in the behavior of water molecules within the nanopore. A larger pore allows a greater number of water molecules to position themselves at the channel opening in a less confined manner. With the reduced energy barrier, water molecules adjacent to the pore face and within it experience an acceleration in transport velocity (transmembrane velocity) in the direction of the applied pressure, resulting in significantly higher flux. This velocity phenomenon and layering within the pores are illustrated in the annualized density and velocity maps.

\begin{figure}[H]
\includegraphics[width=6.4in]{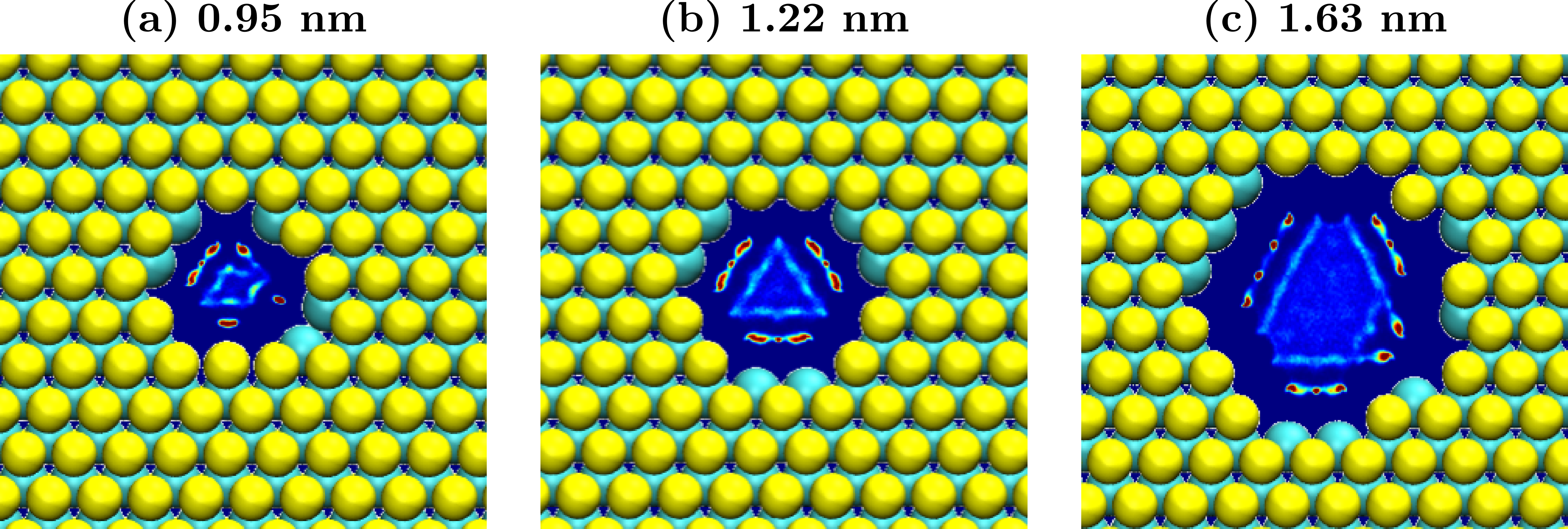}
\caption{Oxygen density maps for membrane under a pressure difference of $\Delta P = 100\; \mathrm{MPa}$. Red regions indicate a high probability of water molecule presence, blue regions indicate low probability, and black regions correspond to areas where no water molecules were detected.}
\label{fig:mapa}
\end{figure}

As can be seen in Figure \ref{fig:mapa}, which shows the density maps of water molecules within each MoS$_{2}$ pore studied, increasing pore diameter also influences how water molecules organize and move through the channel. In very narrow pores, such as those with a diameter of 0.92 nm, water molecules can be forced to pass in a single file or in highly ordered hydrogen-bonded chains. Although this confinement can, in some cases, generate a high velocity for the confined chain due to the high pressure concentrated in the small orifice, the accommodation capacity is limited. With increasing pore diameter (e.g., for diameters of 1.22 and 1.63 nm), the transport channel can accommodate more water molecules, and consequently, this leads to an increase in molecular density, which ultimately dominates the effect and results in a net increase in water flux \cite{suk2013molecular,10.1038/ncomms9616,gao2019water}. An important factor observed in all pores and clearly demonstrated in the density maps is the influence of the atomic termination at the pore edge, which plays a crucial role and is the main mechanism that modulates water flow. The preferential attraction of water molecules to Mo atoms compared to S atoms, and its effect on flow, can be explained by the intrinsic polarity and hydrophilicity/hydrophobicity of these sites. The explanation lies in the fact that in molybdenum disulfide, Mo is less electronegative than S. Consequently, the Mo atoms carry a partial positive charge. The exposure of these Mo sites at the nanopore edge creates highly hydrophilic sites, meaning they attract more water molecules \cite{del2023ultrafast,hong2014ultrafast}. As is known, water is a polar molecule with a significant dipole moment. The oxygen atom carries a partial negative charge, while the hydrogen atoms carry partial positive charges. The oxygen in the water molecule is strongly attracted to the Mo sites at the pore edge, resulting in a favorable and stronger electrostatic interaction compared to the S sites. The S atoms are more electronegative and carry a partial negative charge. When exposed at the pore edge, these sites are less likely to interact with the O in the water and can, in some configurations, act as more hydrophobic (water-repellent) surfaces compared to the Mo border. The electrostatic repulsion between similar partial negative charges hinders the entry and organization of water molecules. Although these factors influence water flow through MoS$_{2}$ pores, in general, what we can observe is that the mobility of water molecules flowing in the pores is highly governed by the pore size and the applied pressure gradient, leaving the interaction between the attraction or repulsion of water molecules with the MoS$_{2}$ atoms in the pores to secondary importance \cite{del2023ultrafast,hong2014ultrafast}.

\begin{figure}[H]
\centering
\includegraphics[width=4.4in]{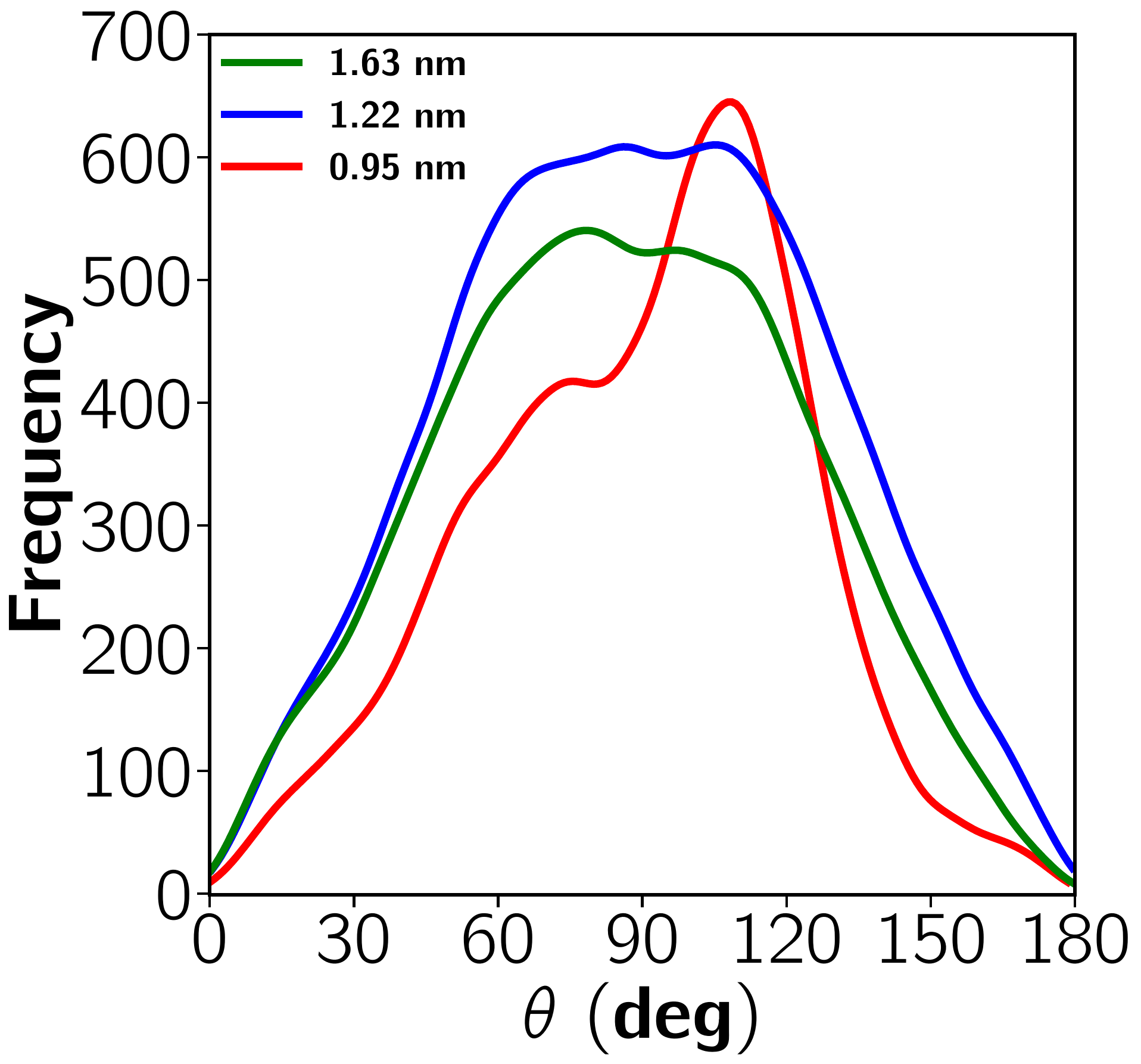}
\caption{Dipole angle distribution of water molecules along the x-direction for MoS$_{2}$  nanopores with diameters of 0.95 nm (red), 1.22 nm (blue), and 1.63 nm (green) under a pressure gradient of $\Delta P = 100\; \mathrm{MPa}$}
\label{fig:dipole}
\end{figure}

The distribution of dipole angles along the x-direction for pores of 0.95 nm, 1.22 nm, and 1.63 nm under a pressure gradient of $\Delta P = 100\; \mathrm{MPa}$ is shown in Fig. \ref{fig:dipole}. For the 1.22 nm and 1.63 nm pores, dipole orientations appear broadly dispersed and randomly distributed, reflecting the greater geometric symmetry and spatial freedom within the channel. This freedom allows water molecules to adopt multiple orientations without a dominant directional preference. In contrast, the 0.95 nm pore imposes significant confinement, which restricts molecular rotation and induces a more ordered dipole alignment. This alignment is driven by both spatial constraints and asymmetric surface interactions at the pore edge.
The asymmetry is from the atomic structure of MoS$_{2}$, where Mo and S atoms are distributed unevenly along the pore edge. In many cases, the pore opening exposes predominantly Mo atoms on one side and S atoms on the opposite side, generating an asymmetric electrostatic field along the pore axis. Mo atoms, carrying partial positive charges, strongly attract the negatively charged oxygen atoms of water. Conversely, S atoms, being more electronegative and also partially negative, tend to repel oxygen and interact weakly with the hydrogen atoms.
This chemical and charge distribution at the pore edge leads to a preferential dipole orientation: oxygen atoms point toward Mo sites, while hydrogen atoms orient toward S atoms. The result is a directional dipole alignment along the pore axis, which is clearly visible in the density maps (Fig. \ref{fig:mapa}) and is consistent with the observed flow behavior. This pattern is especially pronounced in the 0.95 nm pore, where confinement amplifies edge interactions and limits rotational freedom, promoting highly ordered molecular organization.

\begin{figure}[H]
\centering
\includegraphics[width=6.4in]{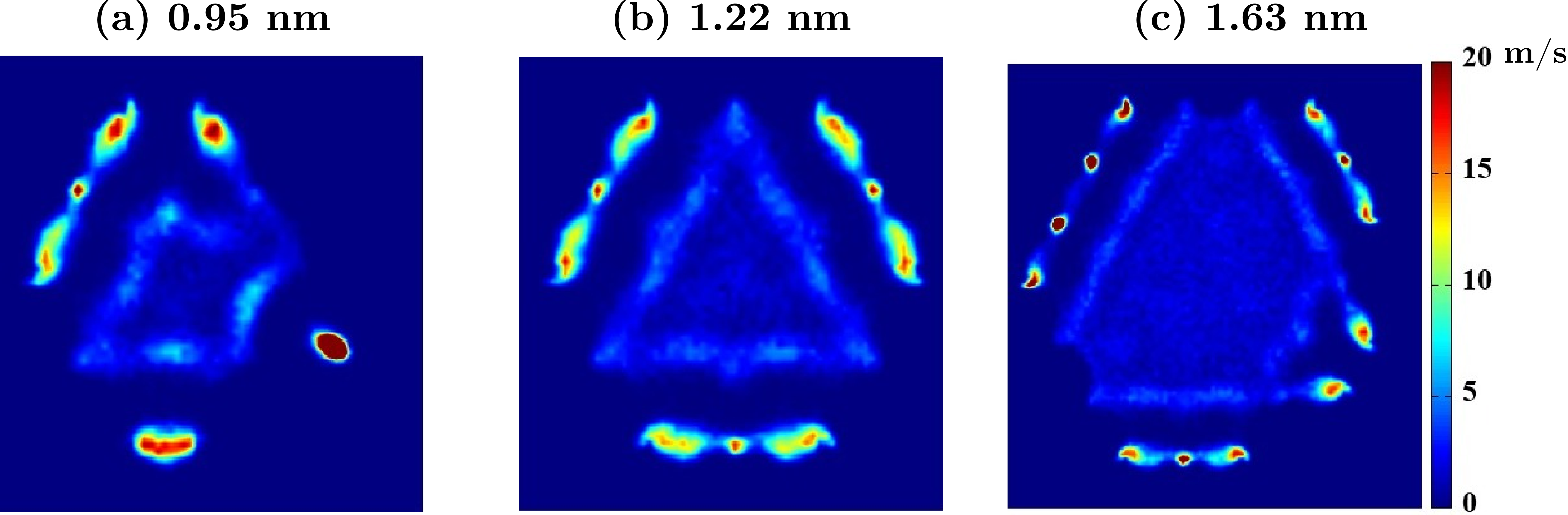}
\caption{Axial velocity profiles of water molecules along the z-direction for membranes under a pressure difference of $\Delta P = 100\; \mathrm{MPa}$.}
\label{fig:v}
\end{figure}

The analysis of axial velocity profiles of water molecules, Fig.\ref{fig:v}, shows that, regardless of pore diameter, the highest velocities are consistently located near the edges of the channel. This behavior contrasts with classical laminar flow, where maximum velocity typically occurs at the center. In MoS$_{2}$ nanopores, this inversion is attributed to strong electrostatic interactions between water molecules and the exposed molybdenum atoms at the pore edge, which act as highly hydrophilic sites. As shown in the density maps (Fig.\ref{fig:mapa}), there is a significant accumulation of water molecules near the pore edges, forming structured layers that promote coordinated flow. Additionally, the preferential dipole alignment observed in narrow pores (Fig. \ref{fig:dipole}) contributes to the stability of molecular trajectories, reducing fluctuations and enabling high velocities along the edges. This peripheral flow pattern is a distinctive feature of water transport in MoS$_{2}$ and may be leveraged to enhance membrane performance in filtration and molecular separation applications. Directly, the speed of water molecules is higher in Mo-rich regions because these areas create a low-friction (high hydrodynamic) flow surface, almost like a molecular ice rink.

\begin{figure}[H]
\includegraphics[width=6.4in]{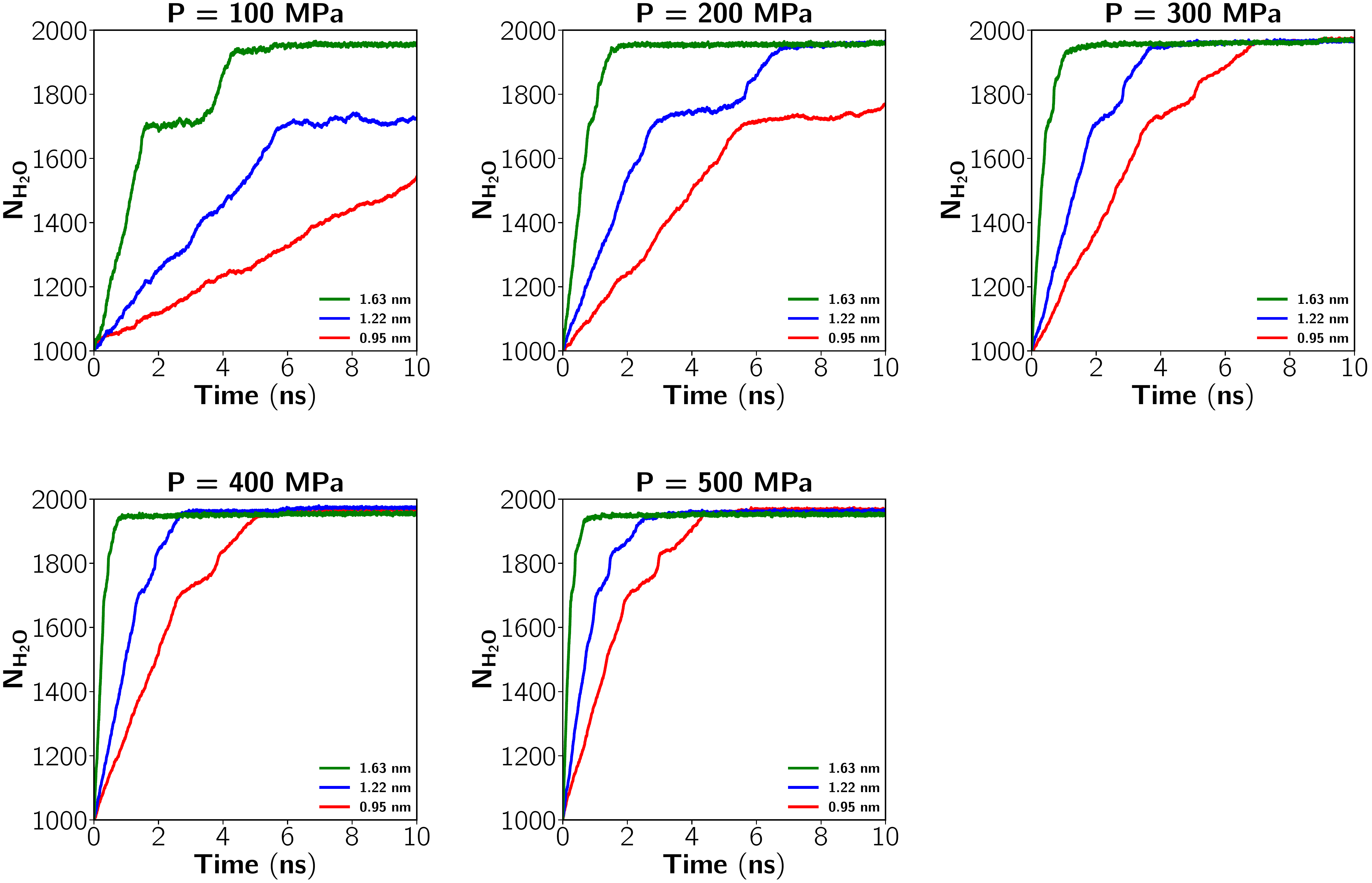}
\caption{Accumulated count of water molecules transported through MoS$_{2}$ nanopores over time for diameters of 0.95 nm (red), 1.22 nm (blue), and 1.63 nm (green).}
\label{fig:counting}
\end{figure}

The accumulated count of water molecules over time, Fig.\ref{fig:counting},offers a clear and intuitive way to understand how efficiently each pore transports water. What we see is straightforward: the larger the pore, the more water gets through - and faster. The 1.63 nm pore leads the way, followed by the 1.22 nm and then the narrower 0.95 nm channel. This isn’t just about size; it’s about how water molecules arrange themselves inside the channel. As shown in the density maps (Fig. \ref{fig:mapa}), wider pores give water molecules room to organize into layered structures, especially near the edges — the same regions where velocity peaks (Fig. \ref{fig:v}). This layered flow, combined with edge acceleration, enhances transport efficiency. In narrower pores, dipole alignment (Fig. \ref{fig:dipole}) helps stabilize the flow, but the tight geometry limits how many molecules can pass through at once. Therefore, the water molecule count reflects the interplay between pore geometry, transport dynamics, and molecular organization, reinforcing pore diameter and symmetry as the dominant factor in flow efficiency.

\section{Conclusions}

In summary, water flows through MoS$_{2}$ nanopores much more efficiently than through graphene pores of similar size. As the pore diameter increases, so does the water flux - not just because there's more space, but because the molecules themselves behave differently. In wider pores, water molecules can spread out, organize into layers, and move with less resistance, especially near the edges of the channel. The density and velocity maps indicate that these edges aren't just boundaries, but active zones. Water tends to accumulate and accelerate along the pore edges, where molybdenum atoms create highly attractive, hydrophilic sites. This edge-driven flow diverges from the classic concept of laminar movement, reminding us that at the nanoscale, atomic details are crucial. The dipole orientation results reinforce this: in narrower pores, water molecules align their dipoles in response to confinement and the asymmetric distribution of Mo and S atoms, creating a directional flow that's both ordered and efficient. Altogether, these findings highlight how pore geometry, edge chemistry, and molecular orientation collectively shape water transport in MoS$_{2}$ membranes. Understanding these interactions helps explain why these systems are so permeable and opens the door to designing smarter, more effective nanofluidic and filtration technologies using transition metal dichalcogenides.

\section{Acknowledgement}

This work is funded by the Brazilian scientific agencies Fundação de Amparo à Pesquisa do Estado da Bahia (FAPESB), Conselho Nacional de Desenvolvimento Científico e Tecnológico (CNPq), and the Brazilian Institute of Science and Technology (INCT) in Carbon Nanomaterials, with collaboration and computational support from Universidade Federal da Bahia (UFBA), and  Universidade Federal de Minas Gerais (UFMG). In addition, the authors acknowledge the National Laboratory for Scientific Computing (LNCC/MCTI, Brazil) for providing HPC resources on the SDumont supercomputer, which contributed to the research results reported in this paper. URL: http://sdumont.lncc.br. Finally,  EEM appreciates Edital PRPPG 010/2024 Programa de Apoio a Jovens Professores(as)/Pesquisadores(as) Doutores(as) - JOVEMPESQ Project 24460. Finally, the authors agree on the PIBIC-UFBA.

\bibliography{apssamp}

\end{document}


\preprint{APS/123-QED}

\title{Material Supplementary:  \\
Exploration of Novel Foam Structures of Carbon,
Boron Nitride and Boron Carbon Nitride: Stability and Electronic Properties}

\author{Juliana Gonçalves}
\thanks{These authors equally contributed to this work.}
\affiliation{Instituto de Física, Universidade Federal da Bahia, Campus Universitário de Ondina, Salvador 40210-340, BA, Brazil}

\author{Elizane E. de Moraes}
\thanks{These authors equally contributed to this work.}
\affiliation{Instituto de Física, Universidade Federal da Bahia, Campus Universitário de Ondina, Salvador 40210-340, BA, Brazil}
\email{elizane.fisica@gmail.com}
\author{Hélio Chacham}
\affiliation{Departamento de Física, ICEX, Universidade Federal de Minas Gerais, CP 702, Belo Horizonte 30123-970, MG, Brazil}
\author{Ronaldo J. C. Batista}
\affiliation{Departamento de Física, Universidade Federal de Ouro Preto, Campus Morro do Cruzeiro, Ouro Preto 35400-000, MG, Brazil}

\date{\today}

\maketitle

\section{Foam structure}
Figures \ref{fig:s1} (a)-(h) show the perspective geometry of the carbon (column 1), BN (column 2), and BCN (columns 3 and 4) configuration foams derived from the structural forms of graphene, diamond, and carbon nanotubes. 
\begin{figure}[h!]
\includegraphics[width=0.8\linewidth]{SI1.pdf}
\caption{Computational unit cell models of the foams carbon, BN, and BCN foam. Carbon, nitrogen,
and boron atoms are grey, blue, and pink, respectively.}
\label{fig:s1}
\end{figure}

\section{BCN planar}
As illustrated in Figure \ref{s2}, the BCN planar structure stands out for its superior energetic stability compared to other geometries proposed \cite{saini2020bcn}. 
\begin{figure}[h!]
\includegraphics[width=1.0\linewidth]{BCNplanar.jpeg}
\caption{BCN planar structure geometries proposed by \cite{saini2020bcn}.}
\label{s2}
\end{figure}

\bibliography{apssamp}